# Breaking Through the Plasma Wavelength Barrier to Extend the Transparency Range of Ultrathin Indium Tin Oxide Films into the Far Infrared

Ran Bi[1], Chuantao Zheng[1], William W. Yu[3], Weitao Zheng [2,*], and Dingdi Wang[1,*]

1. *State Key Laboratory of Integrated Optoelectronics, College of Electronic Science and Engineering, Jilin University, Changchun 130012, China*
2. *Key Laboratory of Automobile Materials, College of Material Science and Engineering, Jilin University, Changchun 130012, China*
3. *School of Chemistry and Chemical Engineering, Shandong University, Jinan 250100, China*

## Abstract

Indium tin oxide (ITO) film, which is the most commonly used transparent conductive film (TCF), has traditionally been believed to be transparent in the visible spectrum but to reflect infrared (IR) light beyond the plasma wavelength ($\lambda_p$). However, our theoretical analysis challenges this notion by demonstrating that an ultrathin ITO TCF that is thinner than the light's penetration depth, can overcome the transmission barrier at $\lambda_p$. To validate the theoretical modeling, we have successfully fabricated ITO films that, despite having $\lambda_p \approx 1$ μm, remain transparent from 400 nm to 20 μm. This represents the broadest transparency range ever reported for any $In_2O_3$-based TCF. The 10-nm-thick ITO TCFs have high visible transmittance (91.0% at 550 nm), low resistivity ($5 \times 10^{-4}$ Ω·cm), and good IR transmittance (averaging 60% over 1.35–18.35 μm). Their IR transparency facilitates radiative cooling of the underlying circuitry. When an operational resistor is enclosed by commercial ITO TCFs that are 140 nm thick, its temperature increases. However, using 10-nm-thick ITO TCFs instead of the commercial ones can completely avoid this temperature rise. Moreover, attaching a silver grid to a 10-nm-thick ITO TCF can reduce the effective sheet resistance to ~10 Ω/□ at the expense of only ~3% transmittance. This development paves the way for large-scale applications that require low sheet resistance and far-IR transparency.

---

* Email: wtzheng@jlu.edu.cn (W. Zheng); wangdingdi@gmail.com (D. Wang)

# 1. Introduction

Transparent conductive films (TCFs) have become indispensable in our daily lives, finding use in touchscreens, flat-panel displays, light-emitting diodes, and solar cells.[1, 2] These films are both electrically conductive and optically transparent in the visible spectrum. However, due to the increasing importance of infrared (IR) sensors in optical communications and thermography,[3-7] the development of IR TCFs is necessary for electromagnetic interference shielding.[8] Mid-infrared TCFs can also serve as transparent electrodes in thermophotovoltaic devices.[9] Additionally, there is a growing need for versatile TCFs that remain transparent across a broad range of wavelengths, from visible to far-IR spectrums. If indium tin oxide (ITO), the most commonly used TCF, allows thermal radiation from internal circuitry to pass through, the radiative cooling effect can lower the circuitry's temperature. Since ITO is integrated into touchscreens and flat-panel displays, our smartphones and tablets can operate at lower temperatures, which enhances their performance and prolongs their battery life. This research aims to expand ITO's transparency range beyond the visible spectrum to the far IR (>15 µm), an area that has received little attention.

Several attempts have been made to extend the transparency range of TCFs into the far-IR region.[10-12] TCFs made of carbon nanotubes have a transparency range of 400 nm to 22 µm,[10] while few-layer graphene films are transparent from 400 nm to over 200 µm.[11, 12] However, the manufacture of these innovative materials is time-consuming and costly, making mass production challenging and limiting their commercialization. Therefore, this study aims to explore the feasibility of extending the transparency range of ITO to the far-IR region, which is a more practical option.

In this paper, we demonstrate that the IR transparency of a TCF is affected not only by its plasma frequency but also by its thickness, which has not received much attention in previous literature.[13, 14] Our theoretical analysis suggests that an ITO TCF with a thickness of 10 nm, a carrier density of $1 \times 10^{21}$ cm$^{-3}$, and a carrier mobility of 10 cm$^2$V$^-$

$^1 s^{-1}$, is transparent across the 0.2–100 μm spectrum. This has been experimentally validated for the 0.4–20 μm spectrum. Extending the transparency to the far-IR range can result in more efficient radiative cooling of internal circuitry. To demonstrate this, we monitored the temperature of an operational resistor enclosed in 10-nm-thick ITO TCFs and commercially available ITO TCFs (140 nm thick). Our results show that using commercial ITO TCFs led to an increase of ~2 °C in the resistor's temperature (~60 °C), whereas this increase was eliminated when we switched to 10-nm-thick ITO TCFs. Furthermore, we discovered that adding a silver grid onto a 10-nm-thick ITO TCF can lower the effective sheet resistance to ~10 Ω/□ with only a ~3% reduction in transmittance. This technique could potentially enable large-scale applications that require both low sheet resistance and far-IR transparency.

## 2. Limitations of using plasma wavelength to determine the cutoff wavelength

The transparency range of a transparent conductive oxide is determined by its band structure. The transmission of short-wavelength light is cut off by inter-band transition processes, which are governed by the bandgap. Conversely, long-wavelength light is reflected and absorbed through intra-band transition processes. According to the Drude model, the cutoff wavelength for long-wavelength light, which is the wavelength beyond which light is totally reflected, can be described by the plasma wavelength ($\lambda_p$),[15]

$$\lambda_p = \frac{2\pi c_0}{\omega_p} = 2\pi c_0 \sqrt{\frac{\varepsilon_0 \varepsilon_\infty m^*}{N e^2}}, \quad (1)$$

where $\omega_p$ represents the plasma frequency, $c_0$ is the speed of light in a vacuum, $\varepsilon_0$ denotes the vacuum permittivity, $\varepsilon_\infty$ represents the high-frequency permittivity, $m^*$ stands for the effective carrier mass, $N$ denotes the carrier density, and $e$ represents the elementary charge.

Conventional wisdom holds that a TCF loses its transparency when the incident light's wavelength exceeds the plasma wavelength ($\lambda_p$), resulting in total reflection of the light.[1, 2, 15-17] In the case of In$_2$O$_3$-based TCFs, the electron density can be reduced to increase $\lambda_p$ in the near-IR range.[18-22] However, the Mott criterion sets a minimum electron density for each conductor,[23] thereby limiting the range of $\lambda_p$ that can be adjusted to a small extent.[24, 25] In this paper, we challenge conventional wisdom by demonstrating that far-IR transparency can be achieved by selecting an appropriate combination of film thickness, electron density, and mobility for a TCF.

The cutoff wavelength of a TCF is determined by the plasma wavelength $\lambda_p$ (or the corresponding plasma frequency $\omega_p$) if and only if two conditions are met: 1) the carrier scattering frequency ($\gamma$) is relatively small ($\omega_p \gg \gamma$), and 2) the thickness ($t$) of the film is significantly greater than the light's penetration depth $\delta_p$ ($t \gg \delta_p$). In simpler terms, if either or both of these conditions are not fulfilled, a film may remain transparent to far-IR radiation without being affected by $\lambda_p$.

First, let us examine how these two conditions ($\omega_p \gg \gamma$ & $t \gg \delta_p$) result in total reflection when $\omega_p > \omega$ (or $\lambda_p < \lambda$). Since $\omega_p > \omega$ and $\omega_p \gg \gamma$, we have $\omega_p > \omega \gg \gamma$. The complex dielectric function $\hat{\varepsilon} = \varepsilon_r + i\varepsilon_i$ is then evaluated as follows:[15]

$$\varepsilon_r = \varepsilon_\infty \left(1 - \frac{\omega_p^2}{\omega^2 + \gamma^2}\right) < 0, \qquad \text{if } \omega_p > \omega \gg \gamma, \qquad (2)$$

$$\varepsilon_i = \frac{\varepsilon_\infty \omega_p^2 \gamma}{\omega(\omega^2 + \gamma^2)} \approx 0, \qquad \text{if } \omega_p > \omega \gg \gamma. \qquad (3)$$

Here, $\varepsilon_r$ and $\varepsilon_i$ represent the real and imaginary parts of the dielectric function, respectively. $\omega$ denotes the angular frequency of light, and $\gamma$ represents the carrier scattering frequency, which is inversely proportional to the mobility $\mu$ by the relation:[26]

$$\gamma = \frac{e}{\mu m^*}. \qquad (4)$$

Given $\varepsilon_r < 0$ and $\varepsilon_i \approx 0$, as shown in Equations (2) and (3), the complex refractive index $\hat{n} = n - i\kappa$ can be calculated as follows:[27]

$$n = \sqrt{\frac{1}{2}\left(\sqrt{\varepsilon_r^2 + \varepsilon_i^2} + \varepsilon_r\right)} \approx 0, \text{ if } \varepsilon_r < 0 \text{ and } \varepsilon_i \approx 0, \tag{5}$$

$$\kappa = \sqrt{\frac{1}{2}\left(\sqrt{\varepsilon_r^2 + \varepsilon_i^2} - \varepsilon_r\right)} \approx \sqrt{|\varepsilon_r|}, \text{ if } \varepsilon_r < 0 \text{ and } \varepsilon_i \approx 0. \tag{6}$$

Here, $n$ and $\kappa$ represent the real and imaginary parts of the complex refractive index, respectively.

When a film is much thicker than the penetration depth of light, $\delta_p$ ($t \gg \delta_p$), the transmitted light cannot reach and reflect off the bottom surface, as illustrated in Fig. 1a. Since light only reflects off the top surface of the film, the reflectance at normal incidence can be easily calculated using the Fresnel equations:[27]

$$R = \left|\frac{\hat{n} - 1}{\hat{n} + 1}\right|^2 = \frac{(n-1)^2 + \kappa^2}{(n+1)^2 + \kappa^2} \approx 1, \text{ if } \kappa > n \approx 0. \tag{7}$$

When $\omega_p > \omega$ (or $\lambda_p < \lambda$), where $\kappa > n \approx 0$, as indicated by Equations (5) and (6), it follows that $R \approx 1$, resulting in total reflection, as derived from Equation (7). Conversely, when $\omega_p < \omega$ (or $\lambda_p > \lambda$), by following the same procedure, it can be straightforwardly deduced that $\varepsilon_r > 0$, which leads to $n > \kappa \approx 0$, and consequently, $R < 1$.

In summary, the cutoff frequency for long-wavelength light (typically observed in the near-IR region) is given by $\omega = \sqrt{\omega_p^2 - \gamma^2} \approx \omega_p$, as defined by $\varepsilon_r = 0$ (Equation (2)). When $\omega_p > \omega \gg \gamma$ and $t \gg \delta_p$, total reflection ($R \approx 1$) occurs. Research has demonstrated that ITO films can be engineered to exhibit $\varepsilon_r \approx 0$ within the near-IR spectrum. The integration of these films into metasurfaces enables precise control of the phase, amplitude, and polarization of light.[28, 29]

In the following two scenarios, the cutoff frequency defined by $\omega_p$ may not accurately predict the behavior of light:

1. Large $\gamma$ value: When $\gamma$ has a significant value that cannot be ignored, the reflectance for $\lambda \gg \lambda_p$ decreases significantly, making total reflection impossible.

In this case, the term $\varepsilon_i$ given by Equation (3) cannot be ignored, and thus $n \approx \frac{\varepsilon_i}{2\sqrt{|\varepsilon_r|}}$ (derived from Equation (5) assuming $\varepsilon_r < 0$) also cannot be ignored. Consequently, the reflectance $R$ calculated using Equation (7) is less than 1 even if $\omega_p > \omega$. Furthermore, if $\gamma$ is increased to the point where $\gamma > \omega_p$, then $\varepsilon_r = \varepsilon_\infty \left(1 - \frac{\omega_p^2}{\omega^2 + \gamma^2}\right) > 0$ holds true even if $\omega_p > \omega$. This causes $n$ to increase and the reflectance $R$ to decrease further. Therefore, as $\gamma$ of a TCF increases, the reflectance for $\lambda \gg \lambda_p$ decreases, allowing significant transmission to occur.

2. Thin film with $t \ll \delta_p$: When the film thickness ($t$) is much smaller than the penetration depth of light, $\delta_p$, Equation (7) is no longer valid. In this case, the light intensity decays only slightly within the thin film, as shown in Fig. 1b. The light waves are reflected and transmitted multiple times at each surface of the thin film, and the interference between these reflected and transmitted waves can be either constructive or destructive. This interference phenomenon has the potential to increase transmissions while decreasing reflections. As a result, significant transmission of light with a wavelength greater than $\lambda_p$ becomes possible, breaking through the plasma wavelength limit.

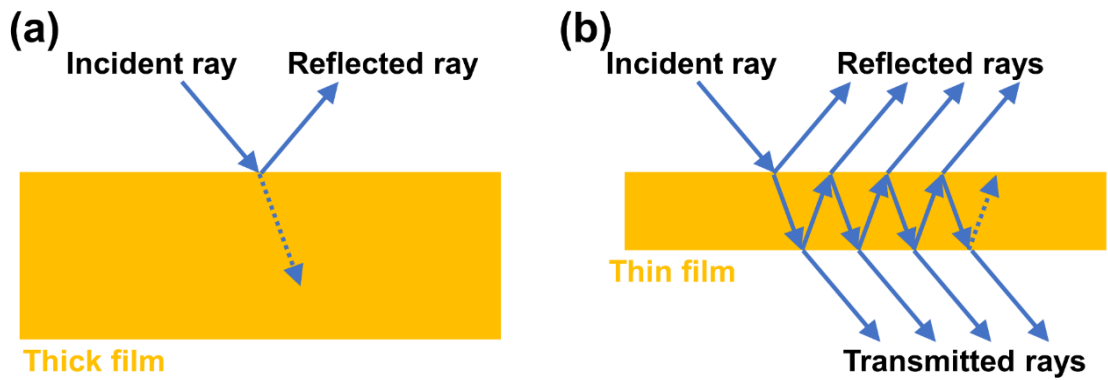

Figure 1. Schematic diagram illustrating the reflection and transmission of light by a film. (a) When the thickness of the film is significantly larger than the penetration depth of light, reflectance calculations only consider the top surface, as the transmitted light does not reach or reflect off the bottom surface. (b) In contrast, when the film is much thinner than the penetration depth of light, multiple reflections and transmissions occur at each surface. These reflected and transmitted light waves can either interfere destructively or constructively, potentially resulting in reduced light reflection from the film and increased light transmission through it.

It is widely recognized that ultrathin metal films can exhibit transparency in the visible range,[2, 30] surpassing the typical limitations imposed by plasma wavelengths observed in the ultraviolet region for thick metals.[2] However, it is important to note that ultrathin metal films maintain their transparency only within the visible and near-IR ranges; they become opaque when the wavelength exceeds 2 μm.[31-35] In contrast, our research focuses primarily on achieving transparency in the far-IR range for ITO films, which distinguishes it from previous studies on ultrathin metal films. The following theoretical modeling example demonstrates how the optical responses of an ITO TCF are influenced by the carrier scattering frequency ($\gamma$) and the film thickness ($t$).

## 3. Theoretical modeling of ultrathin ITO films exhibiting transparency across the visible to far-infrared spectrum

The optical properties of a film were calculated using the transfer-matrix method (see Methods),[36] which considers multiple reflections and transmissions of light at each surface. Figure 2a shows the absorption, reflection, and transmission spectra of an ITO TCF with a thickness of 100 nm, as calculated using this method. The figure illustrates the typical behavior of an ITO TCF, where the film is transparent in the visible spectrum but nearly opaque in the IR spectrum.

In Fig. 2b, an ITO TCF with a mobility of 2 $cm^2V^{-1}s^{-1}$ exhibits remarkable behavior. Compared to the film depicted in Fig. 2a, only mobility is reduced, while all other parameters remain unchanged. When $\lambda > \lambda_p$ = 1.34 μm, both transmittance and absorptance increase compared to those in Fig. 2a, while reflectance drops to 14% and remains at this level over a wide IR range up to 100 μm. Two factors contribute to the reduction in IR reflectance: first, the film depicted in Fig. 2b has a large $\gamma$ ($\gamma/\omega_p$ = 1.6); second, as mobility decreases, the penetration depth increases. The effect of film thickness on the optical properties of a TCF will be further discussed below.

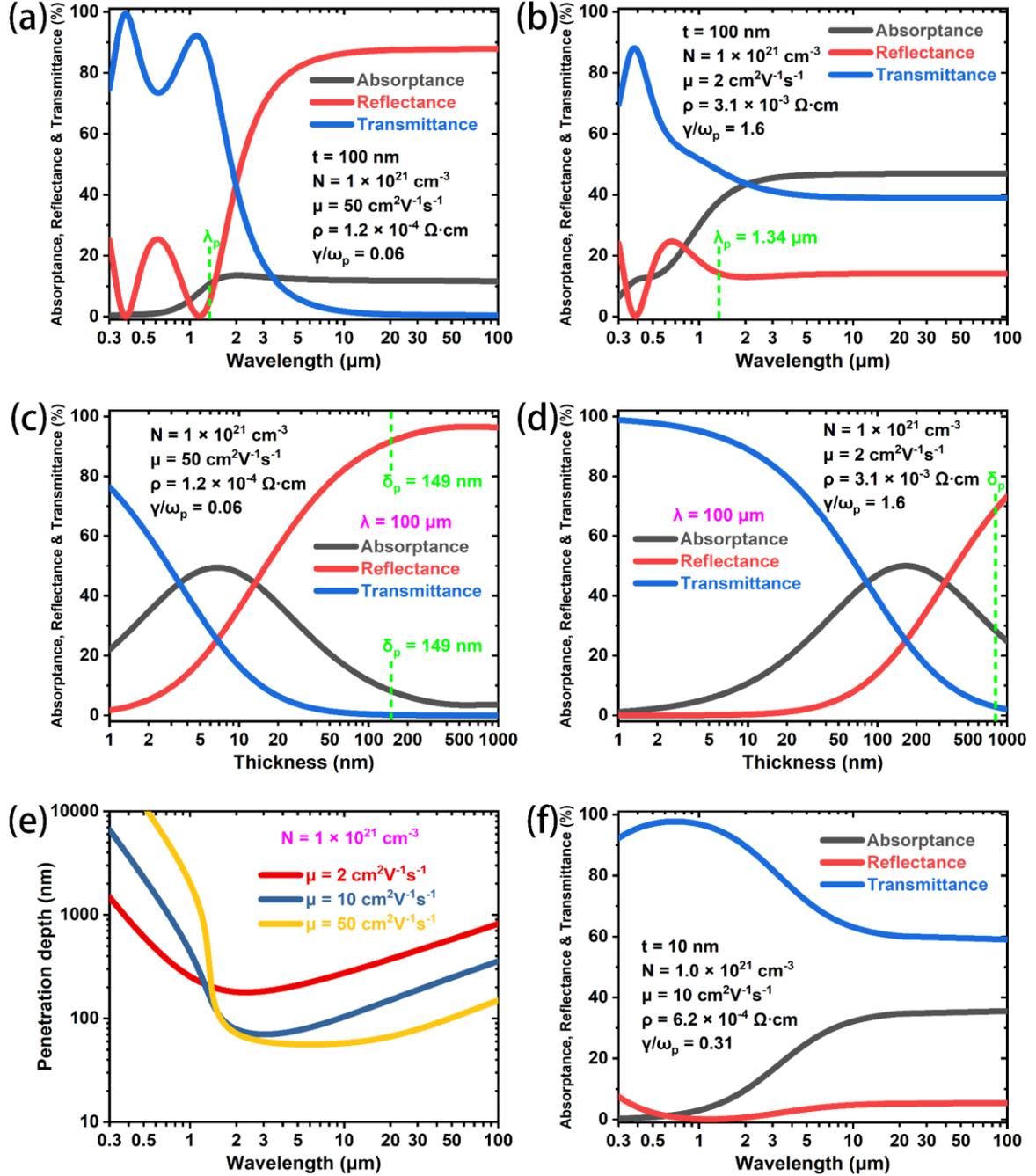

Figure 2. Theoretical optical responses of an ITO TCF suspended in air. (a, b) Calculated absorption, reflection, and transmission spectra for a 100-nm-thick ITO TCF with a mobility of 50 cm²V⁻¹s⁻¹ (a) or 2 cm²V⁻¹s⁻¹ (b) at normal incidence. These spectra demonstrate that reducing mobility can improve the TCF's transparency in the far-IR region. (c, d) Absorptance, reflectance, and transmittance of 100 μm IR light calculated as a function of thickness for an ITO TCF with a mobility of 50 cm²V⁻¹s⁻¹ (c) or 2 cm²V⁻¹s⁻¹ (d). These curves demonstrate that reducing the film thickness can significantly increase the TCF's transparency in the far-IR region. (e) Penetration depth calculated as a function of light wavelength for an ITO TCF with a mobility of 2, 10 or 50 cm²V⁻¹s⁻¹. (f) Calculated absorption, reflection, and transmission spectra for a 10-nm-thick ITO TCF with a mobility of 10 cm²V⁻¹s⁻¹. The electron density is kept constant at 1 × 10²¹ cm⁻³ throughout all calculations.

The absorptance, reflectance, and transmittance of 100 μm IR light, each of which serves as a proxy for the corresponding long-wavelength limit, were calculated as functions of film thickness, as shown in Fig. 2c and 2d. The same electron density and mobility were used to calculate the curves in Fig. 2a and 2c, as well as in Fig. 2b and 2d, respectively. The curves in Fig. 2c indicate that reducing the film thickness to less than 149 nm (the penetration depth of 100 μm IR light) lowers reflectance and increases both transmittance and absorptance. As the film thickness is reduced further ($t$ < 6 nm), absorptance begins to decrease. The three curves in Fig. 2d show a similar trend to those in Fig. 2c, but they are shifted to the right due to the increased penetration depth.

Figure 2e illustrates the penetration depth as a function of light wavelength. A more detailed discussion is available in Part I of the Supplementary Material. The penetration depth rapidly decreases in the near-IR range, dropping below 100 nm when both 1.5 μm < $\lambda$ < 10 μm and 10 $cm^2V^{-1}s^{-1}$ < $\mu$ < 50 $cm^2V^{-1}s^{-1}$. For $\lambda > \lambda_p$ = 1.34 μm, most 100-nm-thick ITO TCFs cannot transmit IR light and instead totally reflect it. To achieve high visible–IR transparency, an ITO TCF must be much thinner than the penetration depth. Our experiments show that ITO can maintain sufficient crystallinity even at a thickness of 10 nm, allowing us to calculate the desired mobility as ~10 $cm^2V^{-1}s^{-1}$ for a transmission limit of 60% (see Fig. S2). The resulting resistivity is ~6.2 × $10^{-4}$ Ω·cm, which is comparable to that of common ITO.

Our goal is to produce an ITO TCF that is 10 nm thick, with an electron density of 1 × $10^{21}$ $cm^{-3}$ and a mobility of 10 $cm^2V^{-1}s^{-1}$. Our calculations indicate that this TCF will exhibit excellent transparency across the visible to far IR spectrum, as shown in Fig. 2f. We plan to produce this ITO TCF using magnetron sputtering.

## 4. Experimental demonstration of ultrathin ITO films exhibiting transparency across the visible to far-infrared spectrum

We fabricated ultrathin ITO TCFs with thicknesses ranging from 5 to 30 nm using magnetron sputtering. The detailed experimental procedures can be found in the Methods section. Firstly, we will showcase the excellent film quality of these ultrathin ITO TCFs, which is critical for achieving both transparency and conductivity. Subsequently, we will analyze and compare the optical and electrical properties of these ultrathin ITO TCFs with the theoretical predictions. Our findings will demonstrate that ultrathin ITO TCFs, particularly those with a thickness of 10 nm, exhibit satisfactory far-IR transparency in addition to high visible transparency and low resistivity.

The film quality of ultrathin ITO TCFs deposited by magnetron sputtering remains high. Figure 3a displays the X-ray diffraction (XRD) pattern of a 10-nm-thick ITO TCF deposited on a fused silica substrate. The diffraction peaks for the (222), (400), (440), and (622) planes of the cubic $In_2O_3$ crystal (PDF 00-006-0416[37]) are evident, confirming that the thin ITO films maintain the $In_2O_3$ crystal structure. Due to lattice expansion in nanocrystals,[38] all peaks shift slightly to lower angles. The broad peak at ~21° can be attributed to the fused silica substrate. Figure 3b shows a scanning electron microscopy (SEM) image of a cross section of a 10-nm-thick ITO film deposited on a Si substrate, which illustrates its uniform thickness.

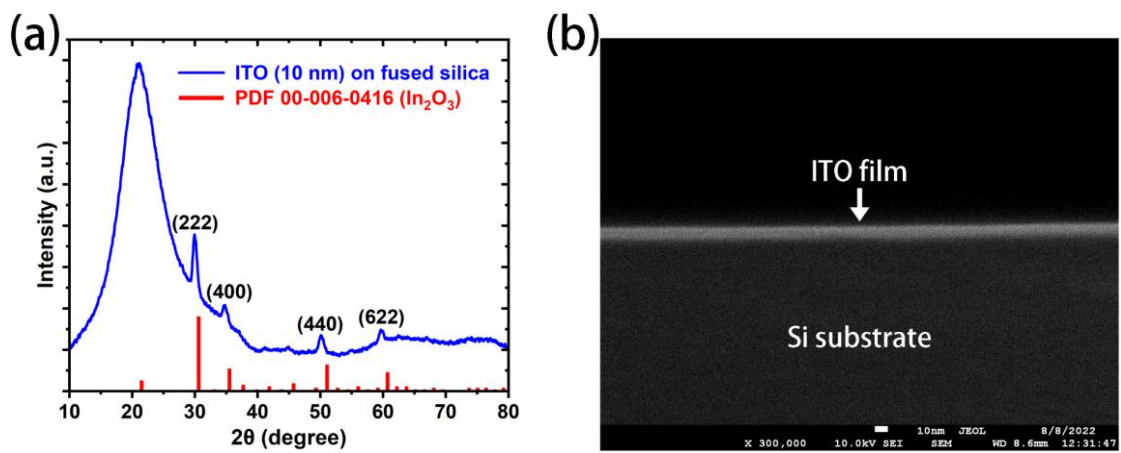

Figure 3. Structure and morphology of 10-nm-thick ITO films. (a) XRD pattern of a 10-nm-thick ITO film deposited on a fused silica substrate, revealing an $In_2O_3$ structure. (b) SEM micrograph of a cross section of a 10-nm-thick ITO film deposited on a Si substrate, demonstrating uniform thickness.

Our next step is to compare the optical properties of ultrathin ITO TCFs with those of a commercial ITO TCF that is 140 nm thick. We have deposited ITO films on two different transparent substrates: fused silica and ZnSe, each with a different transparency range (see Fig. S3). This will enable us to evaluate the transparency of an ITO TCF over a wide range of wavelengths, from 200 nm to 20 µm. The growth of the film on different substrates does not affect its optical properties.

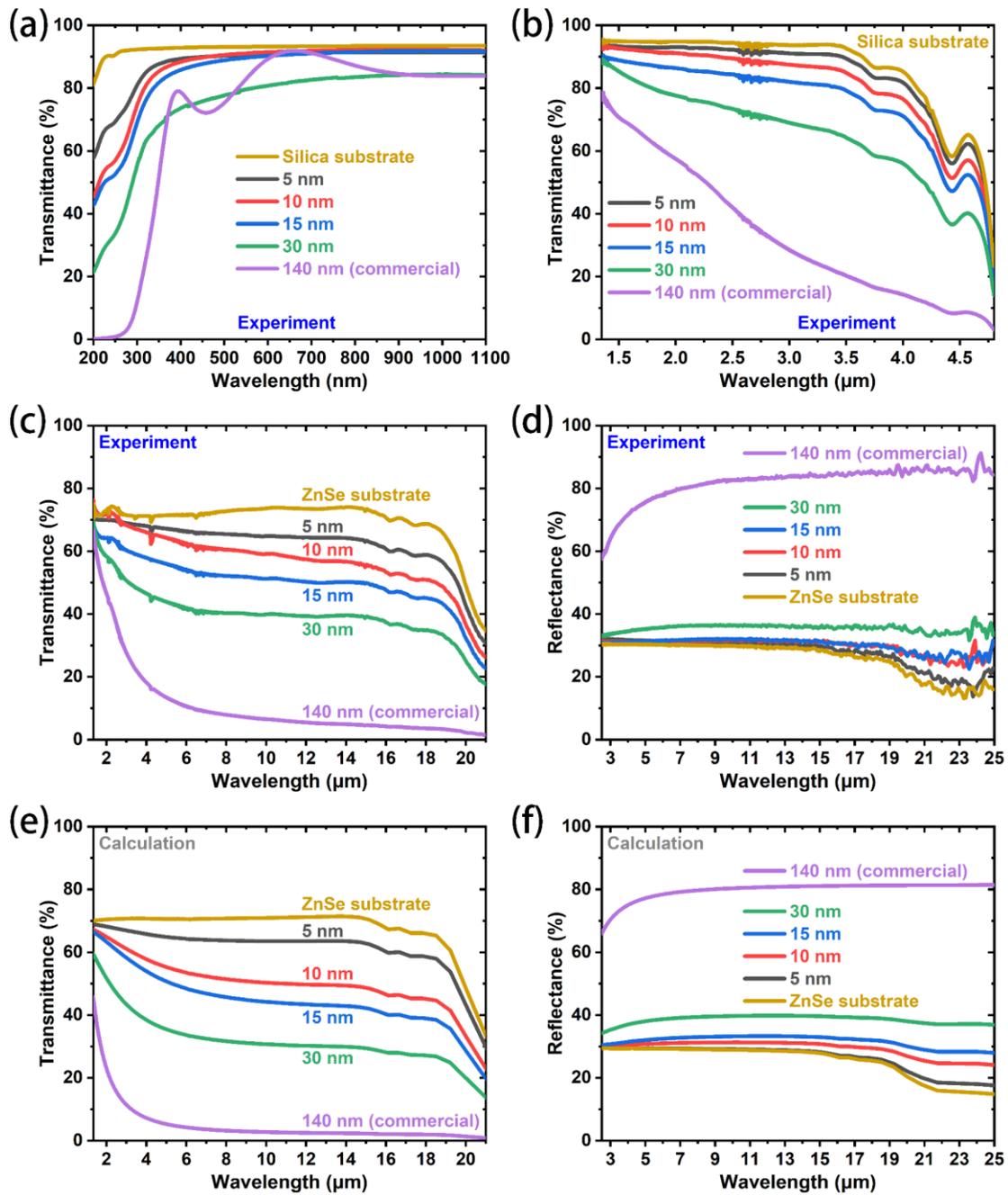

Figure 4. Optical properties of ultrathin ITO films on fused silica and ZnSe substrates. The figure presents experimental normal-incidence transmission spectra of ITO films with varying

thicknesses in three spectral ranges: UV–visible–near-IR (a), near-IR to mid-IR (b), and near-IR to far-IR (c). Panel (d) shows experimental near-normal-incidence reflection spectra (at an 8° angle of incidence) of ITO films deposited on ZnSe substrates. The calculated normal-incidence transmission spectra (e) and near-normal-incidence reflection spectra (f) of ITO films deposited on ZnSe substrates are consistent with the experimental results shown in (c) and (d), respectively.

Figure 4a displays transmission spectra for the 200–1100 nm wavelength range commonly used for TCFs. The figure shows that all TCFs (5–30 nm thick) prepared in-house have high transmittance between 400 and 1100 nm, which is comparable to that of a 140-nm-thick commercial ITO TCF. As shown in Fig. S4, the optical bandgap of a 30-nm-thick ITO film can be determined to be 2.8 eV using a Tauc plot of $(\alpha h\nu)^{1/2}$ vs $h\nu$,[39, 40] where $\alpha$ denotes the absorption coefficient, and $h\nu$ denotes the photon energy. This value is consistent with the bandgap of an $In_2O_3$ crystal, which is ~2.9 eV.[41, 42]

Figure 4b depicts transmission spectra spanning wavelengths from 1.35 to 4.8 μm. The spectra demonstrate that ultrathin TCFs (5–30 nm thick) exhibit high transmittance in the near-IR region. In contrast, a commercially available ITO TCF (140 nm thick) experiences a sharp drop in transmittance, reaching only 20% at 3.5 μm. This comparison clearly demonstrates that reducing the thickness of the ITO film can enhance its transparency in the IR region.

We calculated the transmission spectra of ITO TCFs on ZnSe substrates using the ITO parameters listed in Table 1 and refractive indices of ZnSe from literature sources.[43, 44] The experimental and calculated transmission spectra, spanning wavelengths from 1.35 to 21 μm, are shown in Fig. 4c and Fig. 4e, respectively. The excellent match between the two spectra demonstrates the accuracy of our calculations and validates the previous theoretical modeling.

The high transmittance of ultrathin ITO TCFs can be attributed to their low reflectance. To support this inference, we measured the reflection spectra from 2.5 to 25 μm, as shown in Fig. 4d. The results demonstrate that the reflectance of an ultrathin TCF is low, only slightly higher than that of the ZnSe substrate. In contrast, the reflectance of a commercial ITO TCF (140 nm thick) quickly increases to 80% at 7 μm and remains above 80% from 7 to 25 μm. The calculated reflection spectra in Fig. 4f match the experimental data in Fig. 4d quite well, indicating that thinning the film significantly reduces reflectance and makes ITO TCFs transparent in the far-IR spectrum.

Table 1. Electrical properties of in-house prepared and commercial ITO TCFs

| Substrate | Thickness (nm) | Carrier density ($10^{21}$ cm$^{-3}$) | Carrier mobility (cm$^2$V$^{-1}$s$^{-1}$) | Resistivity ($10^{-4}$ Ω·cm) | Sheet resistance (Ω/□) | Comment |
|---|---|---|---|---|---|---|
| Fused silica | 5 | 0.987 | 16.9 | 3.73 | 746.7 | In-house |
| Fused silica | 10 | 1.13 | 15.7 | 3.51 | 351.2 | In-house |
| Fused silica | 15 | 0.843 | 13.9 | 5.32 | 354.4 | In-house |
| Fused silica | 30 | 1.00 | 11.0 | 5.64 | 188.1 | In-house |
| Fused silica | 140 | 1.65 | 14.5 | 2.61 | 18.64 | Commercial |
| ZnSe | 5 | 0.810 | 8.07 | 9.55 | 1909 | In-house |
| ZnSe | 10 | 1.14 | 10.1 | 5.39 | 539.3 | In-house |
| ZnSe | 15 | 1.03 | 10.9 | 5.56 | 371.0 | In-house |
| ZnSe | 30 | 1.27 | 8.09 | 6.08 | 202.6 | In-house |
| ZnSe | 140 | 1.01 | 20.6 | 3.00 | 21.46 | Commercial |

Table 1 displays the electrical properties of both in-house prepared and commercially available ITO TCFs. There are only minor differences in the electrical properties of films grown on the two different substrates. The deposited ITO follows the island (Volmer–Weber) growth mode,[45] in which islands nucleate and coalesce. Once the ITO film reaches a critical thickness, it becomes continuous, resulting in nearly constant carrier density and mobility, regardless of thickness. Our experimental conditions suggest that the critical thickness of ITO films is ~5 nm, consistent with values reported in the literature.[46,47] Therefore, 10-nm-thick ITO TCFs exhibit good film quality, with electron density and mobility close to our target values of 1 × 10$^{21}$ cm$^{-3}$ and 10 cm$^2$V$^{-1}$s$^{-1}$, respectively.

# 5. Evaluation of the effectiveness of 10-nm-thick ITO TCFs in facilitating radiative cooling

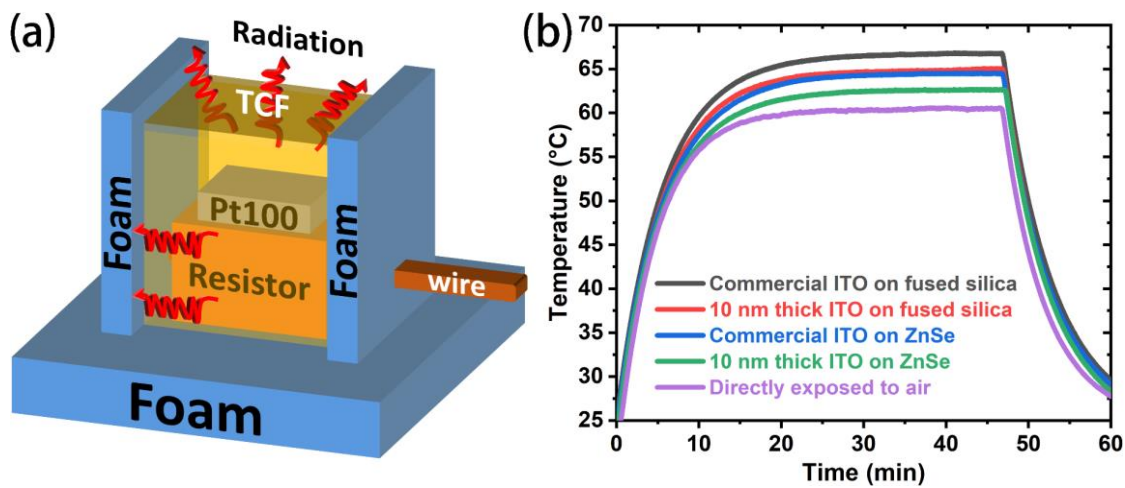

Figure 5. Evaluation of the effectiveness of 10-nm-thick ITO TCFs in facilitating radiative cooling. (a) Schematic illustration of an operational resistor surrounded by three foam walls and three TCF windows. A Pt100 resistance thermometer is attached to the top of the resistor to measure its temperature. (b) Temperature profiles of the operational resistor encased in in-house produced (10 nm thick) or commercial (140 nm thick) ITO TCF windows, or directly exposed to ambient air.

We have developed a model system, as depicted in Fig. 5a, to evaluate the effectiveness of various TCFs in facilitating radiative cooling. In this system, we use an aluminum-housed resistor to represent the internal circuitry of an optoelectronic device, such as a smartphone. To simulate a touchscreen display, we have placed three TCF window pieces that cover 50% of the outer surface surrounding the resistor. The remaining three sides of the resistor are covered with thermal insulation foam.

The temperature of the resistor is measured using a Pt100 resistance thermometer, which is attached to the top of the resistor. Photographs of this model system are displayed in Fig. S5. To maintain a constant heat generation power, a stable electric current is passed through the resistor. Thus, the temperature of the resistor can serve as an indicator of the effectiveness of TCFs in facilitating radiative cooling. According to Wien's displacement law, thermal radiation emitted by an object at a temperature slightly higher than room temperature peaks at ~9 μm (See Fig. S7b).[48] Therefore, a significant amount of the thermal radiation emitted by the resistor can pass through 10-nm-thick ITO TCFs, as shown in Fig. 4c, to assist in cooling the resistor. However, commercial ITO TCFs lose their transparency at ~4 μm, which impedes radiative cooling and causes the resistor's temperature to rise. The Supplementary Material provides a detailed analysis of thermal radiation in this model system.

Table 2. Stable temperatures of the encased resistor

| Sample | $T$ (°C) | $\Delta T$ (°C) |
| --- | --- | --- |
| Commercial ITO TCFs on fused silica substrates | 66.7 | 6.2 |
| 10-nm-thick ITO TCFs on fused silica substrates | 65.0 | 4.5 |
| Fused silica substrates | 65.0 | 4.5 |
| Commercial ITO TCFs on ZnSe substrates | 64.5 | 4.0 |
| 10-nm-thick ITO TCFs on ZnSe substrates | 62.6 | 2.1 |
| ZnSe substrates | 62.5 | 2.0 |
| None | 60.5 | 0 |

Figure 5b depicts temperature profiles of an operational resistor, which is enclosed in either in-house produced (10 nm thick) or commercial (140 nm thick) ITO TCF windows, or exposed directly to air. The figure shows that the resistor's temperature stabilizes after 20 minutes. Table 2 lists the stable temperatures ($T$) of the resistor measured at the 45th minute for various samples, including bare substrates (see Fig. S6). The table also presents the temperature differences ($\Delta T$), with the temperature of the resistor directly exposed to air serving as a reference. The results in Table 2 indicate that depositing 10-nm-thick ITO TCFs on fused silica or ZnSe substrates has almost no impact on the resistor's temperature. However, depositing commercial ITO TCFs on fused silica and ZnSe substrates results in an increase in the resistor's temperature of 1.7°C and 2°C, respectively, compared to bare substrates.

The results above suggest that using 10-nm-thick ITO TCFs can completely prevent the increase in resistor temperature caused by using commercial ITO TCFs. This finding has the potential to optimize TCF design for thermal management in various optoelectronic devices.

## 6. Incorporating a silver grid onto an ultrathin ITO TCF to reduce the effective sheet resistance to 10 Ω/□

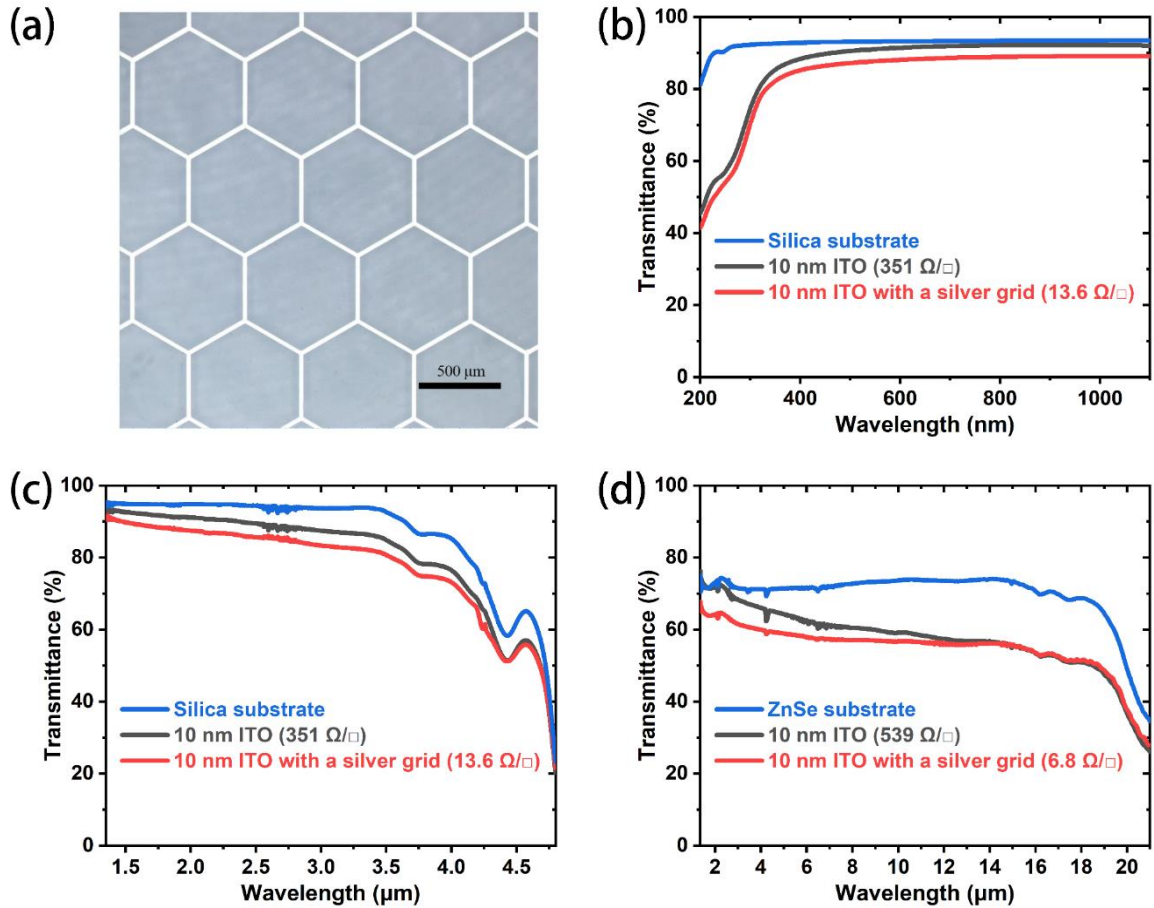

Figure 6. Optical properties of a 10-nm-thick ITO film coated with a silver grid. Panel (a) displays an optical micrograph of a honeycomb-shaped silver grid deposited onto a 10-nm-thick ITO film. The figure displays transmission spectra of the silver grid-coated ITO film with a thickness of 10 nm in three spectral ranges: UV–visible–near-IR (b), near-IR to mid-IR (c), and near-IR to far-IR (d).

We have demonstrated that an ultrathin ITO TCF has high transmittance in the 0.4–20 μm spectrum, and its resistivity is comparable to that of a commercial ITO TCF. However, due to its thinness, the ultrathin ITO TCF has a high sheet resistance in the hundreds of Ω/□, as indicated in Table 1. This level of sheet resistance is suitable for touchscreen applications,[49, 50] but it exceeds the requirements of 100 Ω/□ for display applications and 10 Ω/□ for current-driven applications such as solar cells and light-emitting diodes.[49, 51] To reduce sheet resistance, one potential solution is to fabricate an ultrathin TCF from a more conductive material, which we plan to explore in the future. In this paper, we present a technique for reducing sheet resistance by adding a metal grid to an ultrathin ITO TCF. This method effectively decreases the sheet resistance of a TCF for applications that require full surface conductivity, such as solar cells and light-emitting diodes.[52-56]

Adding a metal grid to a TCF can significantly reduce its effective sheet resistance, making it possible to meet the desired value of 10 Ω/□.[53] The sheet resistance of the metal grid, denoted as $R_s$, can be calculated using the formula $R_s = \frac{\rho}{f_c h}$, where $\rho$ represents the resistivity of the metal, $f_c$ denotes the fraction of the TCF surface covered by the grid's component metal wires, and $h$ represents the grid thickness. For example, if a silver grid with $\rho$ = 1.59 ×10$^{-6}$ Ω·cm is used and $f_c$ = 4%,[57] then $h$ = 39.8 nm can be obtained to achieve $R_s$ = 10 Ω/□. Because the silver grid and the TCF are connected in parallel, their combined sheet resistance will be less than 10 Ω/□.

We have successfully produced a honeycomb-shaped silver grid using the lift-off process, which has a sheet resistance of ~10 Ω/□. The thickness of the silver grid is around 160 nm, which is higher than expected due to the sputtered silver film having a higher resistivity than its bulk value. In Fig. 6a, it can be observed that the sides of a regular hexagon measure 400 μm in length and 14 μm in width. This measurement allows us to calculate $f_c$ = 4%, which determines the theoretical transmittance of the silver grid to be 96%. Figures 6b–6d show the transmission spectra of a silver grid-covered, 10-nm-thick ITO film on a fused silica substrate and a ZnSe substrate. The average transmittance of the hybrid Ag grid/ITO TCF is only slightly lower than that of a 10-nm-thick ITO TCF, decreasing from 91.5% to 88.3% and from 59.8% to 57.1% in the 400–1100 nm and 1.35–18.35 μm spectra, respectively. We believe that the effectiveness of hybrid Ag grid/ITO TCFs in facilitating radiative cooling should be comparable to that of 10-nm-thick ITO TCFs, given their similar IR transmittance. To summarize, adding a silver grid to a 10-nm-thick ITO TCF reduces the effective sheet resistance to ~10 Ω/□ at the expense of ~3% transmittance.

## 7. Conclusions

In conclusion, our theoretical and experimental findings demonstrate that ultrathin ITO TCFs, which are thinner than the light's penetration depth, remain transparent across the visible to far-IR spectrum, regardless of the plasma wavelength. We have successfully produced ITO TCFs that are only 10 nm thick, with a low resistivity of 5 × $10^{-4}$ Ω·cm, and excellent transmittance averaging 91.5% and 59.8% in the 400–1100 nm and 1.35–18.35 μm spectra, respectively. To investigate whether the IR transparency of ultrathin ITO TCFs can facilitate radiative cooling, we monitored the temperature of an operational resistor enclosed in 10-nm-thick or commercial (140 nm thick) ITO TCFs. We observed a ~2 °C increase in the resistor's temperature (~60 °C) when using commercial ITO TCFs, which was completely eliminated when we switched to 10-nm-thick ITO TCFs. Furthermore, by adding a silver grid to a 10-nm-thick ITO TCF, we can reduce the effective sheet resistance to ~10 Ω/□ while only sacrificing ~3% transmittance. This development potentially paves the way for large-scale applications that require both low sheet resistance and far-IR transparency.

## 8. Methods

**8.1 Calculating TCF optical responses**

We utilized an internally developed program, based on the transfer-matrix method,[36] to compute the optical responses of a film. This program requires the film's thickness and the complex refractive index as input parameters. The complex refractive index for ITO can be calculated using equations (1)–(6). ITO exhibits a high-frequency permittivity ($\varepsilon_\infty$) of about 4.[25, 58, 59] At an electron density of 1 × $10^{21}$ cm$^{-3}$, the effective mass ($m^*$) of ITO is considered to be 0.4 $m_e$ (where "$m_e$" denotes the rest mass of an electron), and it increases as the electron density increases.[60-62] The values of the complex refractive index ($\hat{n}$) for the ZnSe substrate are obtained from the literature.[43, 44] The light propagating within the ITO film is considered coherent, while the light within the thick ZnSe substrate is considered incoherent.[63] We conducted consistency checks between our program and an open-source optical simulation software named "Optical".[64] The data depicted in Figs. 2a, 2b, 2f, 3e and 3f, as computed via our internally developed program, are in perfect agreement with those obtained through the "Optical" platform.

**8.2 Preparing ITO film**

ITO films were fabricated using an SP-203 High Vacuum Co-Sputtering System (LJ-UHV Technology Co.). The sputtering process for ITO films was carried out at room temperature on fused quartz substrates (15 × 15 × 0.5 mm) or ZnSe substrates (15 × 15 × 1.5 mm), employing radio frequency magnetron sputtering. A ceramic target of $In_2O_3$ doped with 10 wt% $SnO_2$, measuring 46 mm in diameter, 3 mm in thickness, and with a purity of 99.99%, was utilized. Prior to sputtering, the bare substrates were sequentially cleaned with acetone, ethanol, and deionized water. Before introducing the Ar sputtering gas, the base pressure of the deposition chamber was reduced to less than $8\times10^{-6}$ Torr using a molecular pump. The key sputtering parameters are listed in Table 3. The sputtering time was adjusted to control the thickness of the film. Following the deposition process, the films were annealed in an $N_2$ atmosphere at 500 °C for 30 minutes. Commercial ITO TCFs were obtained from Fuzhou Innovation optoelectronic Technology Co., Ltd.

Table 3. Sputtering conditions for ITO films

| Sputtering Parameters | Values |
|---|---|
| **Base pressure** | $8\times10^{-6}$ Torr |
| **Sputtering pressure** | $1\times10^{-2}$ Torr |
| **Distance between target and substrate** | 14 cm |
| **Ar gas flow rate** | 20 sccm |
| **Substrate temperature** | Room temperature |
| **Sputtering power** | 40 W |

**8.3 Preparing the silver grid**

A silver grid was fabricated on a 10-nm-thick ITO TCF using magnetron sputtering and the lift-off process. Initially, a layer of photoresist was applied to the 10-nm-thick ITO TCF using spin-coating. After that, the photoresist layer was patterned using photolithography, and the pattern was developed to form an inverse honeycomb pattern. Subsequently, a silver film was deposited onto the surface through magnetron sputtering. Finally, the remaining photoresist was washed away, causing the silver that had covered it to be lifted off, leaving behind only the honeycomb-shaped silver grid.

**8.4 Characterizations**

The XRD patterns were recorded using an R-Axis Rapid II X-ray diffractometer (Rigaku Corporation) in a parallel beam geometry with a grazing incidence angle of ~1° and a Cu Kα source (wavelength of 1.54184 Å). The surface morphology of the ITO films was examined using a JSM-7500F field emission scanning electron microscope (JEOL Ltd.). The film thickness was measured with an XP-2 stylus profiler (Ambios Technology Inc.). UV–visible–near-IR transmission spectra were obtained using a UV-1900i spectrophotometer (Shimadzu Corporation) in the wavelength range of 190–1100 nm. IR transmission spectra were obtained using a Nicolet iS50 FTIR spectrometer (Thermo Fisher Scientific Inc.), which has a spectral range of 7400–350 cm$^{-1}$ (1.35–28.57 μm). IR reflection spectra were collected using a Nicolet 6700 FTIR spectrometer (Thermo Fisher Scientific Inc.), with a spectral range of 7400–350 cm$^{-1}$ (1.35–28.57 μm). The van der Pauw method was used to measure the resistivity, Hall mobility, and carrier density of the ITO films with an Accent HL5500PC Hall effect measurement system (Bio-Rad Laboratories Inc.). All measurements were performed at room temperature.

# Supplementary Material

In the supplementary material, we provide Supplementary Fig. S1–S7, along with the relevant discussions.

# Acknowledgment

We would like to express our gratitude to Dr. Haijing Zhang from the Max Planck Institute for Chemical Physics of Solids for her extensive and insightful discussions. This study was funded by the National Natural Science Foundation of China (Grant Nos. 11504124, 61627823) as well as the National Key Research and Development Program of China (Grant No. 2016YFD0700101-03).

# Author Declarations

**Conflict of Interest**

The authors have no conflicts to disclose.

# Supplementary Material

# I. The influence of carrier mobility on light absorption within an ITO film

The carrier mobility plays a significant role in affecting the absorption coefficient ($\alpha$) within the ITO films.

The light intensity within the ITO film undergoes an exponential decay along the propagation direction, such as the x-direction. This phenomenon can be described using the following equation:

$$I(x) = I_0 exp(-\alpha x). \tag{S1}$$

In Equation (S1), the symbol $\alpha$ represents the absorption coefficient, which can be defined as:[1, 2]

$$\alpha = \frac{4\pi\kappa}{\lambda_0}, \tag{S2}$$

where $\lambda_0$ denotes the vacuum wavelength, and $\kappa$ represents the imaginary component of the complex refractive index. The penetration depth of light, denoted by $\delta_p$, is defined as the depth at which the light intensity has decreased to 1/e of its original value just beneath the surface. Consequently, the penetration depth can be calculated using the formula:[1, 2]

$$\delta_p = \frac{1}{\alpha} = \frac{\lambda_0}{4\pi\kappa}. \tag{S3}$$

The impact of carrier mobility on the penetration depth is visually illustrated in Fig. 2e of the main manuscript. Subsequently, we delve into the influence of carrier mobility on the absorption coefficient.

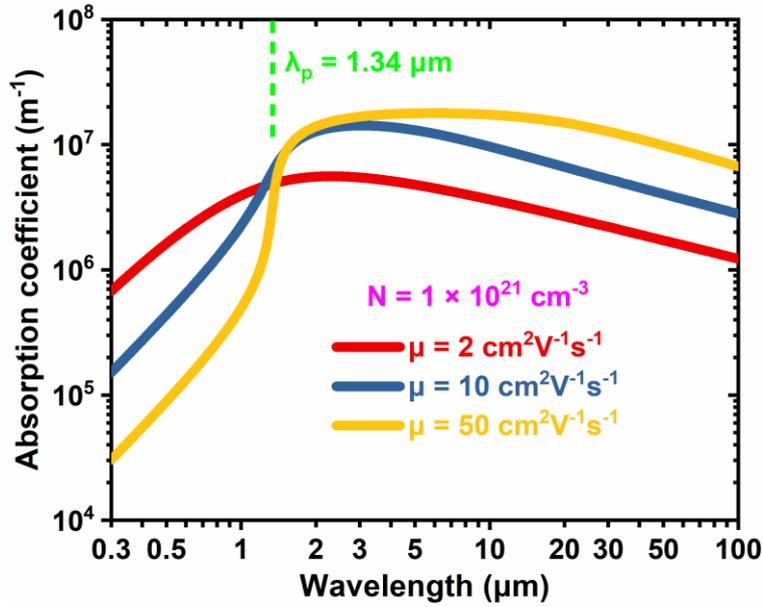

Figure S1. Absorption coefficient calculated as a function of light wavelength for an ITO TCF with a mobility of 2, 10 or 50 $cm^2V^{-1}s^{-1}$.

Figure S1 illustrates the absorption coefficient calculated as a function of light wavelength. The graph illustrates that when $\lambda < \lambda_p$ = 1.34 μm, a decrease in carrier mobility leads to an increase in the absorption coefficient. Conversely, when $\lambda > \lambda_p$ = 1.34 μm, reducing the carrier mobility results in a decrease in the absorption coefficient. Nevertheless, across the entire spectrum, whether $\lambda < \lambda_p$ or $\lambda > \lambda_p$, absorptance increases as carrier mobility decreases, as demonstrated in Fig. 2b. The underlying reason is that absorption is influenced by both the absorption coefficient and the propagation distance, as indicated in Equation (S1), rather than being solely determined by the absorption coefficient itself.

Let us discuss the absorption of IR light by the ITO film. When the carrier mobility is high, for instance, at 50 $cm^2V^{-1}s^{-1}$, the pronounced absorption coefficient in the IR range leads to a rapid decline in light intensity. Consequently, this prevents transmitted light from reaching or reflecting off the bottom surface, as depicted in Fig. 1a. Calculations using the transfer-matrix method reveal that within the IR range, reflectance remains high while absorptance remains low, as illustrated in Fig. 2a. In contrast, when the carrier mobility is low, say 2 $cm^2V^{-1}s^{-1}$, the weak absorption coefficient in the IR range results in only a slight reduction in light intensity within the ITO film. As a result, multiple reflections and transmissions occur at each surface, effectively increasing the distance the light travels and thereby enhancing absorption. In this scenario, the calculation results demonstrate a significant level of absorptance, as depicted in Fig. 2b. This suggests that absorptance depends not only on the absorption coefficient but also on the film's thickness and the refractive indices of the upper and lower layers relative to the ITO film.

## II. Determining the desired carrier mobility for an ITO TCF

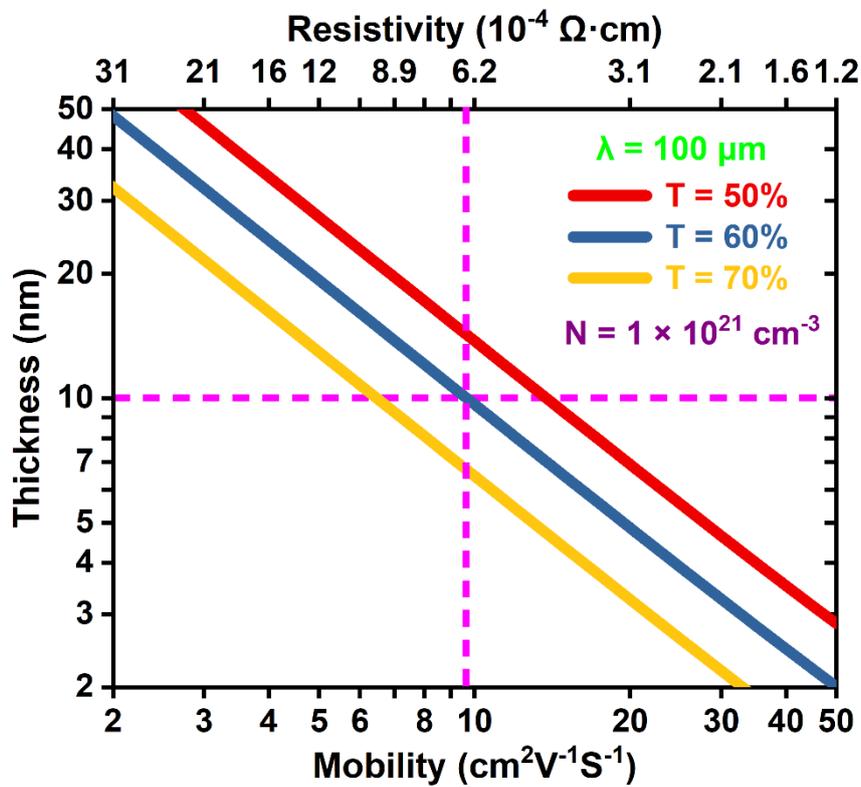

Figure S2. Iso-transmittance curves calculated in the mobility–thickness plane for 100 μm IR light. The electron density is held constant at a value of $1 \times 10^{21}$ cm$^{-3}$ in all calculations.

Our objective is to develop an ITO TCF that exhibits excellent transparency in both the visible and IR spectrum, while maintaining practical resistivity. This goal will be achieved by carefully balancing the film thickness and carrier mobility. As mentioned in the main text, the film should ideally be as thin as possible. Our experimental results have shown that ITO can maintain suitable crystallinity even at a thickness of 10 nm. Therefore, we can determine the desired mobility by drawing a horizontal line on Fig. S2 at $t$ = 10 nm. Figure S2 illustrates three iso-transmittance curves (50%, 60% and 70%) calculated in the thickness–mobility plane for 100 μm IR light. Generally, the far-IR transmittance of 50% is slightly lower, while the transmittance of 70% results in a slightly higher resistivity of $9.6 \times 10^{-4}$ Ω·cm. By examining Fig. S2, we find that the optimal compromise is achieved at the intersection of the iso-transmittance curve of 60% with the iso-thickness line of 10 nm. This intersection determines that the mobility is ~10 $cm^2V^{-1}s^{-1}$ and the resistivity is ~$6.2 \times 10^{-4}$ Ω·cm, which is comparable to that of commonly used ITO.

## III. Transmission spectra of bare substrates

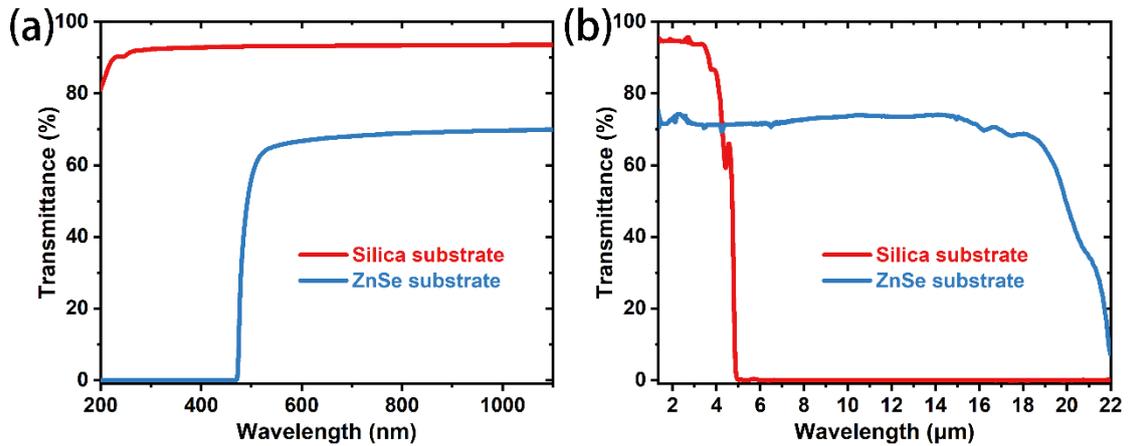

Figure S3. Transmission spectra of bare fused silica and bare ZnSe substrates in the (a) UV–visible–near-IR and (b) near-IR to far-IR wavelength ranges, respectively.

Figure S3 shows the transmission spectra of two different substrates: bare fused silica (15 × 15 × 0.5 mm) and bare ZnSe (15 × 15 × 1.5 mm). These substrates are used to evaluate the overall transmission performance of ITO films deposited on them across the wavelength range of 200 nm to 20 μm. Fused silica demonstrates transparency from 200 nm to 4.8 μm, while ZnSe exhibits transparency from 500 nm to 20 μm.

## IV. Measuring the bandgap of an ITO TCF

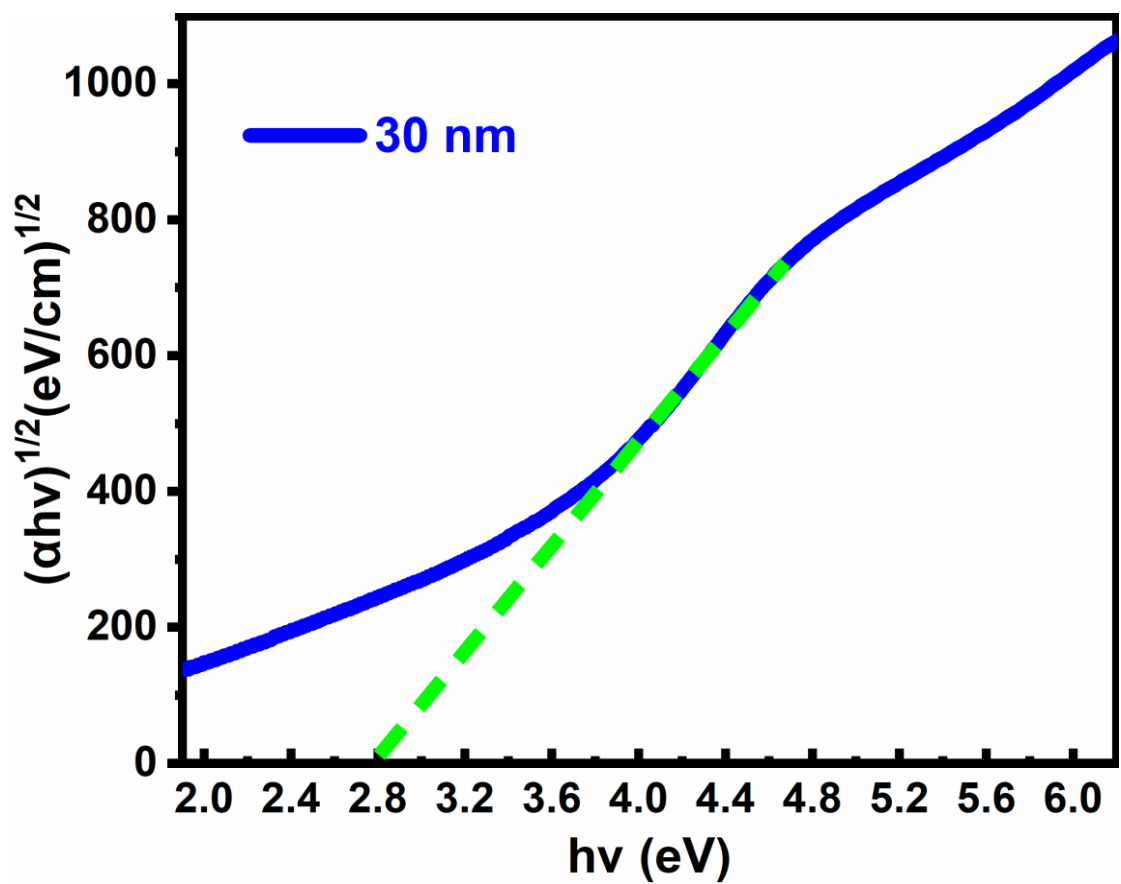

Figure S4. Tauc plot of a 30-nm-thick ITO TCF on a fused silica substrate.

As depicted in Fig. S4, the optical bandgap of a 30-nm-thick ITO film can be determined as 2.8 eV by analyzing a Tauc plot of $(\alpha h\nu)^{1/2}$ versus $h\nu$,[3, 4] where $\alpha$ denotes the absorption coefficient, and $h\nu$ denotes the photon energy. This determined value aligns with the bandgap of an $In_2O_3$ crystal, which is ~2.9 eV.[5, 6]

# V. The model system used to evaluate the effectiveness of various TCFs in facilitating radiative cooling

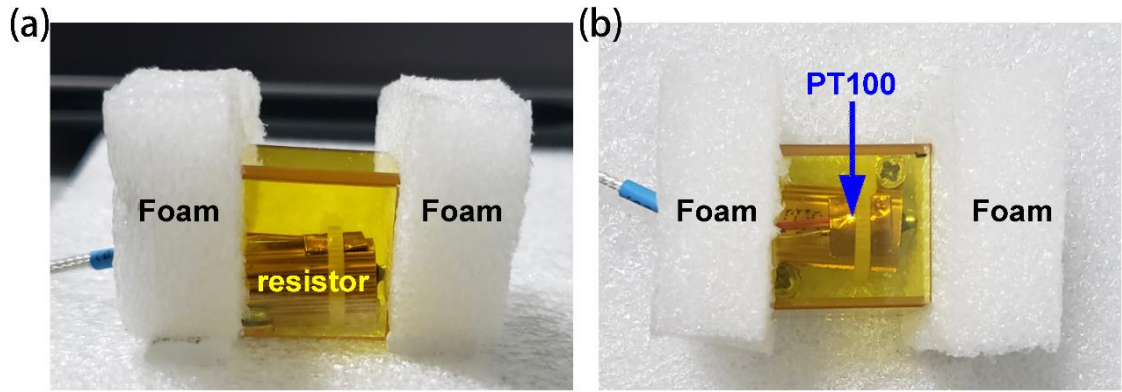

Figure S5. Photographs of the model system used to evaluate the effectiveness of various TCFs in facilitating radiative cooling, captured from the frontal (a) and top (b) views, respectively.

We have developed a model system to evaluate the effectiveness of various TCFs in facilitating radiative cooling. The photographs of the model system are shown in Fig. S5. In this system, we utilize an aluminum-housed resistor with a resistance of 5 Ω to represent the internal circuitry of an optoelectronic device, such as a smartphone. To simulate a touchscreen display, we have placed three TCF window pieces that collectively cover 50% of the outer surface surrounding the resistor. The remaining three sides of the resistor are covered with thermal insulation foam.

The temperature of the resistor is measured using a Pt100 resistance thermometer, which is attached to the top of the resistor (refer to Fig. S5). The temperature is recorded every 0.2 seconds. To ensure consistent heat generation, the electric current flowing through the resistor was maintained at a constant value of 0.359 A for all experiments. Thus, the temperature of the resistor can serve as an indicator of the effectiveness of TCFs in facilitating radiative cooling. Under these conditions, the operating temperature is around 60 °C, which is the typical temperature for an electrical device operating at full capacity. Since the resistor's temperature coefficient of resistance is only 100 ppm/°C, its resistance can be considered stable within the range of 60 °C to 70 °C. All tests were conducted under the same environmental conditions, including a constant room temperature of 23 °C.

# VI. Temperature profiles of an operational resistor encased in uncoated substrates

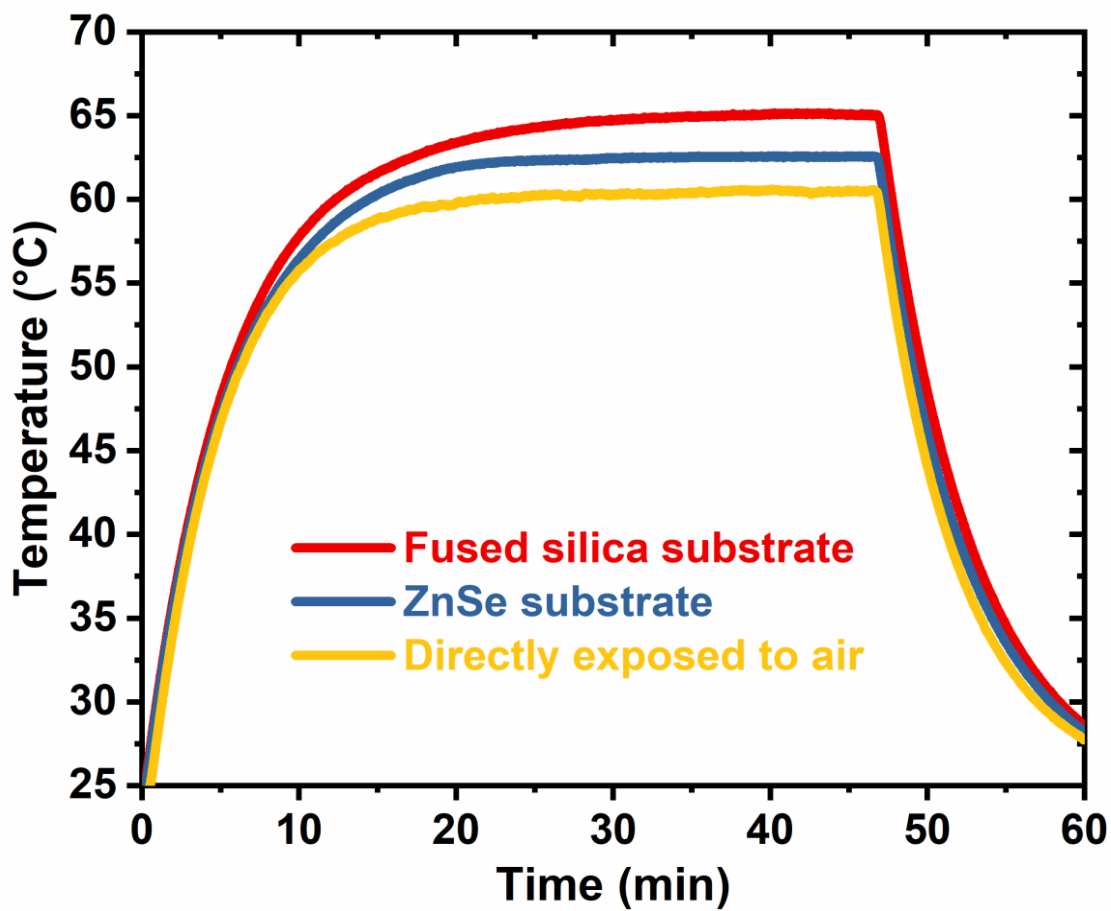

Figure S6. Temperature profiles of a heated resistor enclosed by bare fused silica substrates, bare ZnSe substrates, or directly exposed to ambient air.

Figure S6 depicts temperature profiles of an operational resistor, which is enclosed in either bare fused silica substrates, bare ZnSe substrates, or exposed directly to air. The figure shows that the resistor's temperature stabilizes after 20 minutes. Specifically, the resistor enclosed within bare fused silica substrates maintains a constant temperature of 65.0 °C. However, when bare ZnSe substrates are utilized, the temperature drops to 62.5 °C. This decrease can be attributed to the fact that the resistor's thermal radiation, which peaks at ~9 μm,[7] is strongly absorbed by the fused silica substrate starting at ~4.8 μm, as shown in Fig. S3b, impeding radiative cooling. In contrast, a significant portion of the thermal radiation from the resistor can pass through bare ZnSe substrates since they have a transmittance of ~70% from 500 nm to 18 μm, as illustrated in Fig. S3. When three windows are removed, exposing the resistor directly to air, its temperature drops to 60.5 °C. This temperature represents the lowest achievable temperature for the resistor through radiative cooling, but it is challenging to attain due to the limitations of currently available IR substrates.

## VII. An in-depth analysis of thermal radiation in the model system

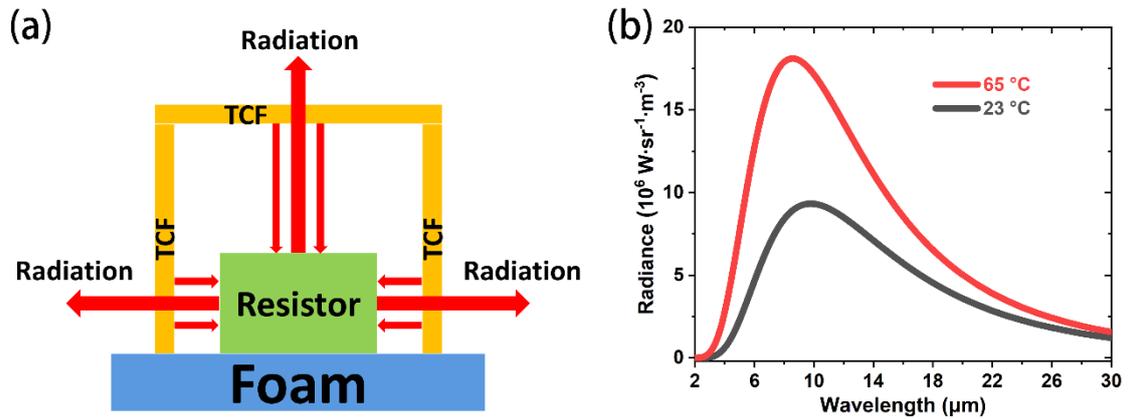

Figure S7. Thermal radiation analysis of the model system used to evaluate the effectiveness of various TCFs in facilitating radiative cooling. (a) A schematic illustrating the thermal radiation emitted by a heated resistor and its surroundings. (b) Spectral distributions of black-body radiation at temperatures of 23 and 65 °C, respectively.

Figure S7a depicts the thermal radiation occurring within the system, which is used to evaluate the effectiveness of various TCFs in facilitating radiative cooling. Thermal radiation is emitted from a heated resistor and its surroundings. The radiative cooling capacity of the heated resistor, referred to as $P_{cool}$, can be described by the following equation:

$$P_{cool} = P_{rad} - P_{abs}. \qquad (S4)$$

In this equation, $P_{rad}$ represents the rate at which thermal energy is emitted by the heated resistor, while $P_{abs}$ represents the rate at which thermal energy is absorbed by the resistor. The resistor absorbs thermal energy emitted by its surroundings, including TCFs and the thermal insulation foam depicted in Fig. S7a. The temperature of these surroundings is approximately equal to the room temperature of 23 °C. In order for $P_{cool}$ to be positive, $P_{rad}$ must exceed $P_{abs}$. According to Planck's radiation law,[8] higher temperatures result in the emission of more thermal energy, as illustrated in Fig. S7b. Consequently, the temperature of the resistor needs to be higher than the room temperature of 23 °C for effective cooling through thermal radiation.

Joule heating can be used to calculate the power of the generated heat as follows:

$$P_J = I^2 R. \qquad (S5)$$

By substituting $I$ = 0.359 A and $R$ = 5 Ω, we can determine $P_J$ to be 0.64 W. Despite the base being made of thermal insulation foam, as shown in Fig. S7a, there is still some heat transfer through the base. This small rate of heat flow is denoted by $P_{con}$ and can be considered constant. Therefore, in a state of thermodynamic equilibrium, $P_{cool}$ remains constant and can be expressed as

$$P_{cool} = P_J - P_{con}. \qquad (S6)$$

By comparing Equations (S4) and (S6), it can be concluded that $P_{rad}$ - $P_{abs}$ is also constant. On the one hand, when $P_{abs}$ is small, $P_{rad}$ is correspondingly small, resulting in a low temperature for the resistor. On the other hand, when $P_{abs}$ is large, $P_{rad}$ should be increased, leading to an elevated temperature for the resistor in accordance with Planck's radiation law.

When we utilize far-IR TCFs, such as 10-nm-thick ITO TCFs, to enclose the resistor, most of the thermal radiation emitted by the resistor (comparing Fig. 4c with Fig. S7b) can pass through the TCFs. Only a small portion of the thermal radiation is reflected (as depicted in Fig. 4d) back to the resistor and gets partially absorbed. However, if we enclose the resistor with commercial ITO TCFs, only 10% of the thermal radiation can pass through the TCFs (comparing Fig. 4c with Fig. S7b), while 80% (as illustrated in Fig. 4d) is reflected back. This results in multiple reflections of IR radiation between the TCFs and the resistor, ultimately increasing the term "$P_{abs}$" in Equation (S4). Since "$P_{cool}$" remains constant, "$P_{rad}$" must be increased. Consequently, it can be concluded that enclosing the resistor with IR-opaque TCFs, such as commercial ITO TCFs, results in an elevated temperature of the resistor.